\documentclass{ifacconf}
\usepackage{amsmath,amssymb,mathtools}
\usepackage{algorithm,algpseudocode}
\usepackage{booktabs}
\usepackage{graphicx}
\usepackage{natbib}
\usepackage[shortlabels]{enumitem}
\usepackage[absolute]{textpos} 

\renewcommand{\qed}{\hfill$\blacksquare$}

\begin{document}
	\begin{frontmatter}
		
		\title{Moving horizon estimation for nonlinear systems with time-varying parameters\thanksref{footnoteinfo}}
		
		\thanks[footnoteinfo]{This work was supported by the Deutsche Forschungsgemeinschaft (DFG, German Research Foundation), Project 426459964.}
		
		\author{Julian D. Schiller}, \author{Matthias A. Müller}
		
		\address{Leibniz University Hannover, Institute of Automatic Control\\
			(e-mail: \{schiller,mueller\}@irt.uni-hannover.de)}

		\begin{abstract}                % Abstract of not more than 250 words.
			We propose a moving horizon estimation scheme for estimating the states and time-varying parameters of nonlinear systems.
			We consider the case where observability of the parameters depends on the excitation of the system and may be absent during operation, with the parameter dynamics fulfilling a weak incremental bounded-energy bounded-state property to ensure boundedness of the estimation error (with respect to the disturbance energy).
			The proposed estimation scheme involves a standard quadratic cost function with an adaptive regularization term depending on the current parameter observability.
			We develop robustness guarantees for the overall estimation error that are valid for all times, and that improve the more often the parameters are detected to be observable during operation. The theoretical results are illustrated by a simulation example.
		\end{abstract}
		
		\begin{keyword}
			Moving horizon estimation, state estimation, parameter estimation.
		\end{keyword}
		
	\end{frontmatter}
	%===============================================================================
	
	\begin{textblock*}{\textwidth}(1.5cm,29.3cm)
		\small{
			© 2024 The Authors. This work has been accepted to IFAC for publication under a Creative Commons Licence CC BY-NC-ND.\\
			10.1016/j.ifacol.2024.09.053
		}
	\end{textblock*}
	
	\section{Introduction}\label{sec:intro}
	
	Reconstructing the state of a dynamical system based on input/output data is an important problem of high practical relevance. This becomes especially challenging in the presence of general nonlinear systems subject to disturbances and noisy output measurements.
	Even if the general system structure is known, the model parameters are often uncertain and/or fluctuate during operation, e.g., due to heat production, mechanical wear, temperature changes or other external influences.
	However, various applications such as (high-performance) control, system monitoring or fault detection often require accurate system models, which demands suitable techniques for online parameter adaptation.
	In this paper, we propose a moving horizon estimation (MHE) scheme to estimate the states and time-varying parameters of general nonlinear systems.
	
	MHE is an optimization-based state estimation technique and hence naturally applicable to nonlinear systems.
	Recently, strong robust stability properties of MHE have been established under a nonlinear detectability condition (incremental input/output-to-state stability, i-IOSS), which is necessary and sufficient for the existence of robustly stable state estimators, cf., e.g., \cite{Allan2021a,Knuefer2023,Schiller2023c}.
	These guarantees, however, do not necessarily hold in the presence of uncertain model parameters.
	To address such a case, \cite{Alessandri2012} proposed a min-max MHE scheme, which yields a state estimation error that is robust with respect to the worst-case model uncertainties.
	In order to provide not only state but also parameter estimates, an MHE scheme was proposed by \cite{Sui2011} by treating the (constant) parameters as additional states (with constant dynamics).
	The corresponding stability analysis requires the transformation of the extended system into an observable and an unobservable subsystem, where the lack of excitation is addressed by a suitable regularization term.
	\cite{Schiller2023d} developed an MHE scheme for joint state and (constant) parameter estimation that results in robust global exponential stability of both the state and parameter estimation error in case of a uniform persistence of excitation condition (or practical stability if the latter is not satisfied). 
	Finally, \cite{Muntwiler2023} used a regularization term that depends on the \textit{a priori} estimate of the (constant) uncertain parameters, where practical stability of the state estimation error with respect to the (constant) \textit{a priori} parameter error could be established.
	
	Alternative approaches to joint state and parameter estimation are provided by adaptive observers, which compute state estimates and simultaneously update internal model parameters, see, e.g., \citep{Ioannou2012}.
	The literature is rich in corresponding results for systems in (linear or nonlinear) adaptive observer canonical forms, mostly under the assumption that the unknown parameter is constant, where stability guarantees typically require a uniform persistence of excitation condition, see,~e.g.,~\citep{Farza2009}.
	These approaches usually can also be applied to track (slowly) time-varying parameters if a forgetting factor is used in the design, see, e.g.,~\citep{Ticlea2016}.
	Time-varying parameters are explicitly considered and analyzed in, e.g., \citep{Bastin1988} and \citep{Marino2001}, requiring that the parameter and its time-derivative are globally bounded for all times.
	
	In this paper, we propose an MHE scheme for robustly estimating states and time-varying parameters of nonlinear systems by extending our recent results developed for constant parameters \citep{Schiller2023d}.
	Whereas considering the states to be uniformly detectable, we assume that the observability of the parameters can depend on the excitation of the system and may be absent during operation, with the parameter dynamics satisfying a weak incremental bounded-energy bounded-state property.
	%This generalizes many state and parameter estimation setups; it covers, for instance, parameters that can be described by a random walk models or that are subject to an unknown deterministic disturbance input.
	This avoids the need to \textit{a priori} guarantee an observability property of the parameters (or satisfaction of a corresponding excitation condition), which is usually impossible for general nonlinear systems.
	Instead, the proposed method relies on online monitoring of parameter observability and selects an appropriate regularization term.
	This allows us to develop robustness guarantees for the overall (state and parameter) estimation error that are valid independent of the parameter observability, and which improves the more often the parameters are detected to be observable during operation.
	%	Different to the most recent MHE results (e.g., \cite{Schiller2023c,Knuefer2023,Muntwiler2023,Schiller2023d}), we consider a standard (non-discounted) least-squares type objective.
	%	Following the underlying idea of \citep{Schiller2023d}, we also consider the case that the initial parameter $z_0$ of an unknown sequence $\{z\}$ can sometimes be observable from the measured outputs during operation and sometimes not.

	\textit{Notation} \ \
	We denote the set of integers by $\mathbb{I}$, the set of all integers greater than or equal to $a \in \mathbb{I}$ by $\mathbb{I}_{\geq a}$, and the set of integers in the interval $[a,b]$ for any $a,b\in \mathbb{I}$ by $\mathbb{I}_{[a,b]}$.
	The $n \times n$ identity matrix is denoted by $I_n$ and the $n \times m$ zero matrix by $0_{n\times m}$.
	The weighted Euclidean norm of a vector $x \in \mathbb{R}^n$ with respect to a positive definite matrix $Q=Q^\top$ is defined as $\|x\|_Q=\sqrt{x^\top Q x}$ with $\|x\|=\|x\|_{I_n}$. The minimal and maximal eigenvalues of $Q$ are denoted by $\underline{\lambda}(Q)$ and $\overline{\lambda}(Q)$, respectively.
	For two matrices $A=A^\top$ and $B=B^\top$, we write $A \succeq B$ ($A\succ B$) if $A-B$ is positive semi-definite (positive definite).
	For $A,B\succ0$, the maximum generalized eigenvalue (i.e., the largest scalar $\lambda$ satisfying $\det (A-\lambda B) = 0$) is denoted by $\overline{\lambda}(A,B)$.
	
	\section{Problem setup}
	We consider nonlinear uncertain discrete-time systems that can be decomposed into the following form
	\begin{subequations}
		\label{eq:sys}
		\begin{align}
			x_{t+1}&=f(x_t,z_t,u_t,w_t), \label{eq:sys_x}\\
			z_{t+1}&=g(z_t,u_t,w_t), \label{eq:sys_z}\\
			y_t&=h(x_t,z_t,u_t,w_t), \label{eq:sys_y}
		\end{align}
	\end{subequations}
	with states $x_t\in\mathbb{R}^{n_x}$, time-varying parameters $z_t\in\mathbb{R}^{n_z}$, control input $u_t\in\mathbb{R}^{n_u}$, (unknown) disturbances $w_t\in\mathbb{R}^{n_w}$, noisy output $y_t\in\mathbb{R}^{n_y}$, and time $t\in\mathbb{I}_{\geq 0}$.
	The nonlinear continuous functions $f$, $g$, and $h$ represent the system and parameter dynamics and the output equation, respectively.
	The system description~\eqref{eq:sys} obviously covers standard setups in the context of parameter estimation:
	\begin{itemize}
		\item Constant parameters $z_t=z$ for some $z\in\mathbb{Z}$ for all $t\in\mathbb{I}_{\geq0}$, where $g(\cdot)=z$ in~\eqref{eq:sys_z}.
		\item Time-varying parameters $z_t$, where $g(z,u,w)$ can be used to model known internal dynamics of $z$, $u$ can represent the influence of known inputs (e.g. explicit dependence of $z_t$ on time) and $w$ can represent, e.g., an unknown drift. An important special case is $g(z,u,w) = z + B_\mathrm{z}w$ for some matrix $B_\mathrm{z}$ and $B_\mathrm{z}w$ is an unknown signal.
	\end{itemize}
	
	In the following, we assume that the states $x$ are uniformly detectable (i-IOSS) and the parameters $z$ are non-uniformly observable (in the sense that the observability depends on the excitation of the system and may be absent during operation). These concepts will be formalized in Section~\ref{sec:obs}.

	\begin{rem}
		Note that the separation between \emph{states} and \emph{parameters} in the description~\eqref{eq:sys} is only for ease of presentation.
		In general, $x$ can represent all time-varying quantities of a (possibly extended) system that are uniformly detectable, and $z$ all quantities that are non-uniformly observable.
		This is particularly relevant for systems in which the observability of certain states depends on the excitation (which can then be described by~\eqref{eq:sys_z}), or for systems in which certain parameters prove to be uniformly detectable (which can then be described by~\eqref{eq:sys_x}).
	\end{rem}
	
	In Section~\ref{sec:MHE}, we propose an MHE scheme to compute estimates $\hat{x}_t$ and $\hat{z}_t$ of the unknown states ${x}_t$ and the parameters ${z}_t$ based on the measured input-output data $(u,y)$ while providing robustness guarantees against arbitrary disturbances $w_t$.
	To this end, we assume that the unknown true system trajectory satisfies
	\begin{equation}\label{eq:con_Z}
		(x_t,z_t,u_t,w_t)\in\mathbb{D} ,\ t\in\mathbb{I}_{\geq 0},
	\end{equation}
	where $\mathbb{D}:=\{(x,z,u,w)\in\mathbb{X}\times\mathbb{Z}\times\mathbb{U}\times\mathbb{W}: f(x,z,u,w)\in\mathbb{X},\ g(z,u,w)\in\mathbb{Z},\ h(x,z,u,w)\in\mathbb{Y}\}$ and $\mathbb{X}\subseteq \mathbb{R}^{n_x}, \mathbb{Z}\subseteq \mathbb{R}^{n_z}, \mathbb{U}\subseteq \mathbb{R}^{n_u}, \mathbb{W}\subseteq \mathbb{R}^{n_w}, \mathbb{Y}\subseteq \mathbb{R}^{n_y}$ are some known closed sets.
	The set $\mathbb{D}$ represents the domain of the real system trajectories and is typically determined by the physical nature of the system.
	Incorporating such information in the estimation scheme can often significantly improve the estimation results, cf. \cite[Sec.~4.4]{Rawlings2017}.
	%Such constraints typically arise from the physical nature of the system, e.g., non-negativity of partial pressures, mechanically imposed limits, or known parameter ranges.
	%Using this additional information in the estimation problem can often significantly improve the estimation results, cf. \cite[Sec.~4.4]{Rawlings2017}.
	If no such information is known, they can simply be chosen as $\mathbb{X}=\mathbb{R}^{n_x}$, $\mathbb{Z}=\mathbb{R}^{n_z}$, $\mathbb{U}=\mathbb{R}^{n_u}$, $\mathbb{W}=\mathbb{R}^{n_w}$, $\mathbb{Y}=\mathbb{R}^{n_y}$.	
	
	Since the parameters $z$ are generally unobservable, we require the following property of $g$~\eqref{eq:sys_z} to ensure boundedness of the estimation error.
	%the system~$g$ in~\eqref{eq:sys_z} must not exhibit unstable internal dynamics in order to provide bounded estimation errors---in other words, $g$ must satisfy an incremental uniform bounded input energy bounded state (i-UBEBS) property.}
	\begin{assum}\label{ass:V}
		There exists a continuous function $V:\mathbb{Z}\times\mathbb{Z}\rightarrow \mathbb{R}_{\geq0}$ and matrices $\underline{V},\overline{V}\succ0$, $Q_{\mathrm{v}}\succeq0$ such that
		\begin{align}
			&\|z-\tilde{z}\|_{\underline{V}} \leq V(z,\tilde{z}) \leq \|z-\tilde{z}\|_{\overline{V}},\label{eq:V_bounds}\\[1ex]
			&V(g(z,u,w),g(\tilde{z},u,\tilde{w})) - V(z,\tilde{z}) \leq \|w-\tilde{w}\|_{Q_{\mathrm{v}}} \label{eq:V_dissip}
		\end{align}
		for all $(x,z,u,w),(\tilde{x},\tilde{z},u,\tilde{w}) \in \mathbb{D}$.
	\end{assum}

	Assumption~\ref{ass:V} essentially constitutes an incremental Lyapunov characterization of a uniform bounded-energy bounded-state (UBEBS) stability property.
	This concept was originally introduced by \cite{Angeli2000a} for continuous-time systems and is frequently employed in the context of integral input-to-state stability, see, e.g., \cite{Haimovich2020,Liu2022a}. A (non-incremental) Lyapunov version of UBEBS is also termed \emph{zero-output dissipativity} by~\cite{Angeli2000}.
	Note that Assumption~\ref{ass:V} does \emph{not} impose asymptotic stability of $\|z_t-\tilde{z}_t\|$, which would render our estimator design task trivial.
	Instead, it merely implies that the difference $\|z_t-\tilde{z}_t\|$ is bounded at each time $t\in\mathbb{I}_{\geq0}$ in terms of the initial difference $\|z_0-\tilde{z}_0\|$ and the sum of the input differences $\|w_j-\tilde{w}_j\|, j\in[0,t-1]$, and particularly does not diverge because of unstable internal dynamics.
	While this may seem restrictive at first glance, it is actually very intuitive and in fact necessary to provide bounded estimation errors with respect to the disturbance energy for the case when $z$ is unobservable during operation (which is also covered in our setup).
	Note that for the common special case where $g(z,u,w) = z + B_\mathrm{z}w$, Assumption~\ref{ass:V} can be satisfied by choosing $V(z,\tilde{z})=\|z-\tilde{z}\|$ and $Q_\mathrm{v} = B_\mathrm{z}^\top B_\mathrm{z}$.
	
	%In the following section, we formalize the concepts of nonlinear detectability and observability used in this work.
	
	\section{Detectability and observability concepts}\label{sec:obs}
	We consider the following nonlinear detectability property with respect to the state $x$.
	
	\begin{defn}[i-IOSS Lyapunov function]\label{ass:W}
		\, A continuous function $W:\mathbb{X}\times\mathbb{X}\rightarrow \mathbb{R}_{\geq0}$ is an (exponential) \mbox{i-IOSS} Lyapunov function for the system~\eqref{eq:sys} if there exist matrices $\underline{W},\overline{W},S_{\mathrm{w}},Q_{\mathrm{w}},{R}_{\mathrm{w}}\succ0$ and a constant $\eta_\mathrm{w}\in[0,1)$ such that
		\begin{align}
			&\|x-\tilde{x}\|_{\underline{W}}^2 \leq W(x,\tilde{x}) \leq \|x-\tilde{x}\|_{\overline{W}}^2,\label{eq:W_bounds}\\[1ex]
			&W(f(x,z,u,w),f(\tilde{x},\tilde{z},u,\tilde{w}) \nonumber \\
			&\hspace{3ex} \leq \eta_\mathrm{w} W(x,\tilde{x}) + \|z-\tilde{z}\|_{S_\mathrm{w}}^2 + \|w-\tilde{w}\|_{Q_\mathrm{w}}^2  \nonumber  \\
			& \hspace{3ex} \phantom{\leq} \ + \|h(x,z,u,w)-h(\tilde{x},\tilde{z},u,\tilde{w})\|_{R_\mathrm{w}}^2 \label{eq:W_dissip}
		\end{align}
		for all $(x,z,u,w),(\tilde{x},\tilde{z},u,\tilde{w}) \in \mathbb{D}$.
	\end{defn}

	Definition~\ref{ass:W} is equivalent to the exponential version of the typically employed notion of i-IOSS when interpreting the parameter $z$ as a time-varying exogenous input in~\eqref{eq:sys_x} (see, e.g., \cite{Allan2021,Schiller2023c}).
	This concept became standard as a description of nonlinear detectability in the context of MHE (for state estimation) in recent years, see, e.g., \cite{Allan2021a,Hu2023,Knuefer2023,Rawlings2017,Schiller2023c}.
	We point out that Assumption~\ref{ass:W} is not restrictive; by an extension of the results from~\cite{Allan2021,Knuefer2023}, one can show that this constitutes a necessary and sufficient condition for the existence of robustly globally exponentially stable state estimators if the true parameter $z_t$ can be measured, and for practically stable state estimators with respect to the error $\max_{0\leq j \leq t-1}\|z_j-\tilde{z}_j\|$, where $\tilde{z}_j$ is an \textit{a priori} guess of the unknown parameter $z_j$.
	Note that i-IOSS Lyapunov functions in the sense of Definition~\ref{ass:W} can be numerically computed using linear matrix inequalities and semidefinite programming, cf.~\cite{Schiller2023c,Arezki2023}.

	\begin{defn}[Observable trajectory pairs]\label{def:obs}
		The tuple
		\begin{equation*}
			X{ =} \left(\hspace{-0.5ex}\left\{(x_t,z_t,u_t,w_t)\right\}_{t=0}^{T-1},\left\{(\tilde{x}_t,\tilde{z}_t,u_t,\tilde{w}_t)\right\}_{t=0}^{T-1}\right)\in\mathbb{D}^T\times\mathbb{D}^T
		\end{equation*}
		is an observable trajectory pair of length $T\in\mathbb{I}_{\geq0}$ of the system~\eqref{eq:sys} if $X\in\mathbb{E}_T$, where
		\begin{align}
			\mathbb{E}_T := \Big\{&\left(\left\{(x_t,z_t,u_t,w_t)\right\}_{t=0}^{T-1},\left\{(\tilde{x}_t,\tilde{z}_t,u_t,\tilde{w}_t)\right\}_{t=0}^{T-1}\right)\nonumber \\
			& \in\mathbb{D}^T\times\mathbb{D}^T: \nonumber\\
			&x_{t+1}=f(x_t,z_t,u_t,w_t), \ \tilde{x}_{t+1}=f(\tilde{x}_t,\tilde{z}_t,u_t,\tilde{w}_t),\nonumber\\
			&z_{t+1}=g(z_t,u_t,w_t),\ \tilde{z}_{t+1}=g(\tilde{z}_t,u_t,\tilde{w}_t),\nonumber\\
			&y_t=h(x_t,z_t,u_t,w_t),\ \tilde{y}_t=h(\tilde{x}_t,\tilde{z}_t,u_t,\tilde{w}_t),\nonumber\\	
			&t\in\mathbb{I}_{[0,T-1]},\nonumber\\[-2ex]	
			&\|z_0-\tilde{z}_0\|^2_{S_\mathrm{o}} \leq \eta_\mathrm{o}^T\|x_{0}-\tilde{x}_{0}\|^2_{P_\mathrm{o}}+ \sum_{j={0}}^{T-1}\|w_{j}-\tilde{w}_{j}\|^2_{Q_\mathrm{o}} \nonumber\\[-1ex]
			& \hspace{14.5ex}+ \sum_{j={0}}^{T-1}\|y_j-\tilde{y}_j\|^2_{R_\mathrm{o}} \Big\} \label{eq:E_set}
		\end{align}
		for some $S_\mathrm{o},P_\mathrm{o},Q_\mathrm{o},R_\mathrm{o}\succ0$, and $\eta_\mathrm{o}\in[0,1)$.
	\end{defn}

	An observable pair of trajectories in the sense of Definition~\ref{def:obs} essentially implies a robust $T$-step observability property with respect to the initial condition $z_0$.
	For the special case of $x_0=\tilde{x}_0$ and without disturbances ($w,\tilde{w}\equiv0$), this implies that the initial parameter $z_0$ (and hence the whole sequence $\{z_j\}_{j=0}^T$) of an unknown trajectory can be computed exactly on the basis of a batch of $T$ output measurements.
	A similar observability property has often been used in the literature in the context of state estimation, see, e.g., \cite{Wynn2014,Alessandri2008,Rao2003}.
	The additional terms depending on the differences $x_0-\tilde{x}_0$ and $w_j-\tilde{w}_j$, $j\in\mathbb{I}_{[0,T-1]}$ represent a robust generalization and stem from the fact that we consider an unknown initial state $x_0\neq\tilde{x}_0$ and the presence of unknown disturbances $w\neq\tilde{w}$.
	Note that Definition~\ref{def:obs} does \emph{not} require that all possible pairs of trajectories are element of the set $\mathbb{E}_T$ for all times, which would imply that $z_0$ is uniformly observable for any input sequence.
	Instead, we use Definition~\ref{def:obs} in Section~\ref{sec:MHE} by monitoring if the currently estimated trajectory segments form an observable pair or not and selecting an appropriate regularization term.
	
	\section{Moving horizon estimation}\label{sec:MHE}
	
	In this section, we present the MHE scheme (cf. Sec.~\ref{sec:MHE_design}) and provide a corresponding robust stability analysis (cf.~Sec.~\ref{sec:MHE_analysis}).
	
	\subsection{Design}\label{sec:MHE_design}
	At each time $t\in\mathbb{I}_{\geq 0}$, we consider the measured past input-output sequences of the system~\eqref{eq:sys} within a (finite) moving window of length $N_t = \min\{t, N\}$ for some $N\in\mathbb{I}_{\geq0}$. More precisely, the current estimates $\hat{x}_t$ and $\hat{z}_t$ are obtained by solving the following nonlinear program:
	\begin{subequations}\label{eq:MHE}
		\begin{align}\label{eq:MHE_0}
		&\min_{\hat{x}_{t-N_t|t},\hat{z}_{t-N_t|t},\hat{w}_{\cdot|t}}\  J_t(\hat{x}_{t-N_t|t},\hat{z}_{t-N_t|t},\hat{w}_{\cdot|t}) \\ 
		&\text{s.t. }
		\hat{x}_{j+1|t}=f(\hat{x}_{j|t},\hat{z}_{j|t},u_j,\hat{w}_{j|t}), \ j\in\mathbb{I}_{[t-N_t,t-1]}, \label{eq:MHE_x} \\	
		&\hspace{0.6cm} \hat{z}_{j+1|t}=g(\hat{z}_{j|t},u_j,\hat{w}_{j|t}), \ j\in\mathbb{I}_{[t-N_t,t-1]}, \label{eq:MHE_z} \\	
		&\hspace{0.6cm} \hat{y}_{j|t}=h_{}(\hat{x}_{j|t},\hat{z}_{j|t},u_j,\hat{w}_{j|t}),\ j\in\mathbb{I}_{[t-N_t,t-1]}, \label{eq:MHE_y} \\	
		&\hspace{0.6cm}(\hat{x}_{j|t},\hat{z}_{j|t},u_j,\hat{w}_{j|t})\in\mathbb{D},\ j\in\mathbb{I}_{[t-N_t,t-1]} \label{eq:MHE_con}
		\end{align}
	\end{subequations}
	for all $t\in\mathbb{I}_{\geq0}$. The decision variables $\hat{x}_{t-N_t|t}$, $\hat{z}_{t-N_t|t}$, and the sequence $\hat{w}_{\cdot|t} = \{\hat{w}_{j|t}\}_{j=t-N_t}^{t-1}$ denote the estimates of the states and parameters at the beginning of the horizon and the disturbance sequence over the horizon, respectively, estimated at time $t$.
	These uniquely define a sequence of state and parameter estimates $\{\hat{x}_{j|t}\}_{j=t-N_t}^{t}$ and $\{\hat{z}_{j|t}\}_{j=t-N_t}^{t}$ under \eqref{eq:MHE_x} and \eqref{eq:MHE_z} for a given input sequence $\{u_j\}_{j=t-N_t}^{t-1}$.
	We consider the cost function
	\begin{align}
		&J_t(\hat{x}_{t-N_t|t},\hat{z}_{t-N_t|t},\hat{w}_{\cdot|t}) \nonumber \\ 
		& = 2\gamma(N_t)\|\hat{x}_{t-N_t|t}-\bar{x}_{t-N_t}\|_{\overline{W}}^2   + 2\eta^{N_t}\|\hat{z}_{t-N_t|t}-\bar{z}_{t-N_t}\|_{\overline{V}}^2 \nonumber \\
		& \quad +\sum_{j=1}^{N_t}2\|\hat{w}_{t-j|t}\|_{Q}^2+\|\hat{y}_{t-j|t}-y_{t-j}\|_{R}^2 , \label{eq:MHE_objective}
	\end{align}
	where $\{y_j\}_{j=t-N_t}^{t-1}$ is the measured output sequence of system~\eqref{eq:sys}.
	The prior estimates $\bar{x}_{t-N_t}$ and $\bar{z}_{t-N_t}$ are defined below.
	The cost function parameters are chosen depending on the parameters of Assumption~\ref{ass:V} and Definitions~\ref{ass:W} and~\ref{def:obs}. In particular, we select $Q = Q_w + Q_v + Q_\mathrm{o}$, ${R} = R_\mathrm{w} + R_\mathrm{o}$, and the discount factors
	\begin{align}
		&\gamma(s) = \eta_{\mathrm{w}}^s +\overline{\lambda}(P_\mathrm{o},\overline{W})\eta_{\mathrm{o}}^s,\label{eq:gamma}\\
		&\eta \in (\max\{\eta_{\mathrm{w}},\eta_{\mathrm{o}}\},1). \label{eq:eta_1}
	\end{align}
	Note that tuning the cost function in practice is possible by re-scaling the functions $W,V$, or the parameters of the set $\mathbb{E}_T$, compare also~\cite[Rem.~1]{Schiller2023c}.
	
	The MHE estimates are defined as
	\begin{equation}\label{eq:MHE_estimates}
		\hat{x}_t = \hat{x}^*_{t|t}, \qquad \hat{z}_t =  \hat{z}^*_{t|t},
	\end{equation}
	and the corresponding estimation error as
	\begin{equation}\label{eq:MHE_error}
		e_t^\top
		= \begin{bmatrix}  e_{\mathrm{x},t}^\top, & e_{\mathrm{z},t}^\top	\end{bmatrix}
		= \begin{bmatrix}  (\hat{x}_t-x_t)^\top, & (\hat{z}_t-z_t)^\top	\end{bmatrix}.
	\end{equation}
	
	In the cost function~\eqref{eq:MHE_objective}, we choose the prior estimate $\bar{x}_{t-N_t}=\hat{x}_{t-N_t}$ for all $t\in\mathbb{I}_{\geq0}$.
	The prior estimate $\bar{z}_{t-N_t}$, on the other hand, is adapted to the currently detected level of observability of $\bar{z}_{t-N_t}$.
	To this end, for any time $t\in\mathbb{I}_{\geq 0}$, let
	\begin{equation*}
		X_{t}{:=}\left(\{(\hat{x}_{j|t}^*,\hat{z}_{j|t}^*,u,\hat{w}_{j|t}^*)\}_{j=t-N_t}^{t-1}{,}\{({x}_{j},z_j,u_j,{w}_{j})\}_{j=t-N_t}^{t-1}\right)
	\end{equation*}
	denote the pair of optimal and true trajectories of length $N_t$.
	Then, we choose the prior estimate $\bar{z}_t$ (that will be used in the cost function~\eqref{eq:MHE_objective} in $N$ time steps in the future) according to the following update rule:
	\begin{equation}\label{eq:pbar}
		\bar{z}_{t} =
		\begin{cases}
		\hat{z}_t,& \text{if $t\in\mathbb{I}_{\geq N}$ and $X_t\in\mathbb{E}_T$},\\
		g^{(N_t)}(\bar{z}_{t-N_t}), & \text{otherwise}
		\end{cases}
	\end{equation}
	for $t\in\mathbb{I}_{\geq1}$ and $\bar{z}_0 = \hat{z}_0$, where $g^{(N_t)}(\bar{z}_{t-N_t})$ is the solution of the system \eqref{eq:sys_z} after $N_t$ steps initialized at $\bar{z}_{t-N_t}$ and driven by the inputs $\{u_j\}_{j=t-N_t}^{t-1}$ and $w_j=0$, $j\in\mathbb{I}_{[t-N_t,t-1]}$ for all $t\in\mathbb{I}_{\geq0}$.
	
	\begin{rem}
		The proposed MHE scheme requires monitoring whether $X_t\in\mathbb{E}_T$ currently holds or not. This is, however, non-trivial for general nonlinear systems since the second trajectory of the pair in $X_t$ is the true, unknown system trajectory.
		It can, however, be checked locally by adapting, e.g., the techniques from~\cite{Sui2011} or \cite{Flayac2023} that are based on the construction of nonlinear observability maps using the mean-value theorem evaluated at the estimated trajectory. Alternatively, it is possible to extend the method proposed in Section 5 of \citep{Schiller2023d} (for constant parameters $z_t=z$ for all $t\in\mathbb{I}_{\geq0}$) to the more general setup considered here, which involves the construction of an observability metric based on certain matrix recursions, compare~\cite[Rem.~16, Rem.~17]{Schiller2023d}.
 	\end{rem}

	\subsection{Stability analysis}\label{sec:MHE_analysis}
	
	In this section, we establish fundamental stability properties of the estimation error~\eqref{eq:MHE_error}.
	The general idea of the proof is similar to the one we used in \citep{Schiller2023d}. Technical difficulties due to the fact that the parameter $z_t$ evolves according to the dynamical system~\eqref{eq:sys_z} are addressed by invoking Assumption~\ref{ass:V}.
	%Further challenges have emerged as Assumption~\ref{ass:V} (in particular, the properties~\eqref{eq:V_bounds} and \eqref{eq:V_dissip}) is defined in terms of the Euclidean norm, while the cost function~\eqref{eq:MHE_objective} should remain quadratic to ensure the applicability of a standard least-squares type cost function.
	
	We consider the Lyapunov function candidate
	\begin{equation}\label{eq:candidate}
		\Gamma(c,\hat{x},x,\hat{z},z_{t}) = W(\hat{x},x) + cV(\hat{z},z)^2, \quad c\geq 1.
	\end{equation}
	The following two auxiliary lemmas establish boundedness properties of $\Gamma$ depending on the observability of $z$; the proofs can be found in the Appendix in Sections \ref{sec:App1} and \ref{sec:App2}.
	
	\begin{lem}\label{lem:nonPE}
		Let the system~\eqref{eq:sys} be i-IOSS according to Definition~\ref{ass:W} and Assumption~\ref{ass:V} be satisfied. Consider the MHE scheme~\eqref{eq:MHE} with cost function \eqref{eq:MHE_objective}.
		Assume that $t\in\mathbb{I}_{[0,N-1]}$ or $t\in\mathbb{I}_{\geq N}$ and $X_t\notin\mathbb{E}_N$. Then, the estimates~\eqref{eq:MHE_estimates} satisfy
		\begin{align}
			&\Gamma(1,\hat{x}_t,x_t,\hat{z}_t,z_t)\nonumber\\
			&\leq  \eta^{-N}c_1(1,N_t)\Big((2\eta_\mathrm{w}^{N_t} 	+ 2\gamma(N_t))
			\|\bar{x}_{t-N_t}-{x}_{t-N_t}\|_{\overline{W}}^2 \nonumber\\
			&\quad + 4\eta^{N_t}\|\bar{z}_{t-N_t}-{z}_{t-N_t}\|^2_{\overline{V}}+ 4\sum_{j=1}^{N_t}\|w_{t-j}\|^2_{Q} \Big)\label{eq:lem_nonPE}
		\end{align}
		with
		\begin{equation}\label{eq:c1_s}
			c_1(r,s) =  \max\{r,\overline{\lambda}(S_{\mathrm{w}},\underline{V})\eta_{\mathrm{w}}^{-1}\}(s+1)\frac{1-\eta^{s+1}}{1-\eta}.
		\end{equation}
	\end{lem}

	\begin{lem}\label{lem:PE}
		Let the system~\eqref{eq:sys} be i-IOSS according to Definition~\ref{ass:W} and Assumption~\ref{ass:V} be satisfied. Consider the MHE scheme~\eqref{eq:MHE} with cost function \eqref{eq:MHE_objective}.
		Assume that at some time $\mathbb{I}_{\geq N}$, $X_t\in\mathbb{E}_N$. Then, the corresponding estimates~\eqref{eq:MHE_estimates} satisfy
		\begin{align}	
			&\Gamma(c,\hat{x}_t,x_t,\hat{z}_t,z_t)\nonumber\\
			& \leq 
			\mu\Gamma(1,\bar{x}_{t-N},x_{t-N},\bar{z}_{t-N},z_{t-N})\nonumber\\
			& \quad + 4c_{1}(c,N)\max\{1,\overline{\lambda}(\overline{V},S_o)\}\sum_{j={1}}^{N}\|w_{t-j}\|_{Q}^2\label{eq:lem_PE}
		\end{align}
		with
		\begin{align}
			\mu :=&\ c_1(c,N)\max\{1,\overline{\lambda}(\overline{V},S_o)\}\nonumber\\
			&\cdot 
			\max\{4\overline{\lambda}(\overline{W},\underline{W})\gamma(N),2\overline{\lambda}(\overline{V},\underline{V})\eta^N\} \label{eq:mu}
		\end{align}
		for all $c\geq 1$.
	\end{lem}
	
	Now, let	
	\begin{align}
		\rho :=
		\eta^{-N}c_1(1,N)(2\eta_\mathrm{w}^{N} + 2\gamma(N))
		\overline{\lambda}(\overline{W},\underline{W})\label{eq:rho}
	\end{align}	
	and $c$ be such that
	\begin{equation}\label{eq:c_def}
		c = \frac{8c_{1}(1,N)\overline{\lambda}(\overline{V},\underline{V})}{1-\rho}+2
	\end{equation}
	with $c_{1}$ from~\eqref{eq:c1_s}.
	The robustness guarantees for the MHE scheme presented in Section~\ref{sec:MHE_design} require that $\eta,N$ are chosen such that
	\begin{align}
		&\max\{\mu,\rho\} <1.\label{eq:contraction}
	\end{align}
	Note that this is always possible, since $\eta$ satisfies condition~\eqref{eq:eta_1} and $c_1$ is dominated by a factor that exponentially decreases with $N$.

	In the following, we divide each time $t\in\mathbb{I}_{\geq0}$ into time intervals (horizons) of length $N$ and the remainder $l=t-\lfloor{t/N}\rfloor N$. We now proceed with some definitions that will help us to formalize our statements.
	First, we define the set of time instances at which the corresponding MHE optimization problem was solved involving an observable trajectory pair as
	\begin{equation}
		\mathcal{T}_t := \left\{\tau\in\mathbb{I}_{[N,t]} : t-\tau-N\left\lfloor\frac{t-\tau}{N}\right\rfloor=0, X_{\tau}\in\mathbb{E}_{N}\right\}.
	\end{equation} 
	By $k\in\mathbb{I}_{\geq0}$, we denote the cardinality of $\mathcal{T}_t$, i.e., $k:= |\mathcal{T}_t|$.
	Furthermore, we define the sequence of times $\{t_m\}_{m=1}^{k+1}$ by $t_1 := \max\{\tau\in\mathcal{T}_t\}$,
	\begin{align*}
		t_{m+1} & := \max\{\tau\in\mathcal{T}_t:\tau<t_m\}, \ m\in\mathbb{I}_{[1,k-1]},
	\end{align*}
	and $t_{k+1} := l$.
	
	The following lemma establishes boundedness of the estimation error on $[t_1,t]$ and a decrease in Lyapunov coordinates on each interval $[t_{m+1},t_m]$ for all $m\in\mathbb{I}_{[1,k]}$.
	\begin{lem}\label{lem:bounds}
		Let the system~\eqref{eq:sys} be i-IOSS according to Definition~\ref{ass:W} and Assumption~\ref{ass:V} be satisfied. Consider the MHE scheme~\eqref{eq:MHE} with cost function \eqref{eq:MHE_objective}.
		Suppose that $\eta$ and $N$ satisfy~\eqref{eq:contraction}.
		Then, there exists $c_{\mathrm{Q},1},c_{\mathrm{Q},2}>0$ such that the estimates~\eqref{eq:MHE_estimates} satisfy the following bounds:\\
		\begin{align}
			\Gamma(1,\hat{x}_{t},x_{t},\hat{z}_{t},z_t) \leq&\ \Gamma(c,\hat{x}_{t_1},x_{t_1},\bar{z}_{t_1},z_{t_1}) \nonumber\\&\ + c_{\mathrm{Q},1}\left(\sum_{r=t_1}^{t-1}\|w_{r}\|_{Q}\right)^2\label{eq:proof_P1_res}
		\end{align}
		and
		\begin{align}
			\hspace{-1.25ex}\Gamma(c,\hat{x}_{t_m},x_{t_m},\hat{z}_{t_m},z_{t_m})
			\leq&\ \mu\Gamma(c,\hat{x}_{t_{m+1}},x_{t_{m+1}},\bar{z}_{t_{m+1}},z_{t_{m+1}})\nonumber\\
			& +  c_{\mathrm{Q},2}\left(\sum_{r={t_{m+1}}}^{t_m-1}\|w_{r}\|_{Q}\right)^2 \label{eq:proof_P2_res}
		\end{align}
		for all $t\in\mathbb{I}_{\geq0}$, all $m \in \mathbb{I}_{[1,k]}$, all $\hat{x}_0,x_0\in\mathbb{X}$, all $\hat{z}_0,z_0\in\mathbb{Z}$, and every sequence $\{u_r\}_{r=0}^{\infty}\in\mathbb{U}^\infty$ and $\{w_r\}_{r=0}^{\infty}\in\mathbb{W}^\infty$, where we recall that $t_{k+1}=l$ for any $k\in\mathbb{I}_{\geq0}$.
	\end{lem}

	\begin{pf}
		We start with the bound~\eqref{eq:proof_P1_res}. 
		Let $j$ denote the number of unobservable horizons that occurred between time $t$ and $t_1$, i.e., $j := (t-t_1)/N$. First, assume that $j\in\mathbb{I}_{\geq1}$.
		From Lemma~\ref{lem:nonPE}, the fact that $\|\bar{x}_{t-N}-x_{t-N}\|_{\overline{W}}^2 \leq \overline{\lambda}(\overline{W},\underline{W})W(\bar{x}_{t-N},x_{t-N})$ by~\eqref{eq:W_bounds}, and the definition of $\rho$ from~\eqref{eq:rho}, we obtain
		\begin{align*}
			&\Gamma(1,\hat{x}_{t},x_{t},\hat{z}_{t},z_t)\\
			&\leq 
			\rho W(\bar{x}_{t-N},x_{t-N})+ c_1' \|\bar{z}_{t-N}-z_{t-N}\|^2_{\overline{V}}  \\
			& \quad + c_1'\eta^{-N}\sum_{r={t-N}}^{t-1}\|w_{r}\|_{Q}^2 \nonumber
		\end{align*}
		with $c_1'=4c_1(1,N)$, and where $\rho$ satisfies $\rho<1$ due to satisfaction of~\eqref{eq:contraction}.
		Note that $\bar{x}_{t-N}=\hat{x}_{t-N}$ and $\bar{z}_{t-N}$ satisfies the update rule~\eqref{eq:pbar}, which implies that $\bar{z}_{t-qN} = g^{((j-q)N)}(\bar{z}_{t-jN})$ for all $q\in[0,j]$ and $t-jN=t_1$.
		From Assumption~\ref{ass:V} and Jensen's inequality, it further follows that
		\begin{align}
			V(\bar{z}_{t-qN},z_{t-qN})^2\leq 2V(\bar{z}_{t_1},z_{t_1})^2 + 2\left(\sum_{r=t_1}^{t-qN-1}\|w_r\|_{Q_\mathrm{v}}\right)^2. \label{eq:proof_V}
		\end{align}
		Combined, we can write that
		\begin{align}
			&\Gamma(1,\hat{x}_{t},x_{t},\hat{z}_{t},z_t)\nonumber\\
			&\leq 
			\rho W(\bar{x}_{t-N},x_{t-N})+ 2c_1'\overline{\lambda}(\overline{V},\underline{V})V(\bar{z}_{t_1},z_{t_1})^2\nonumber \\
			&\quad + c_1'\max\{2\overline{\lambda}(\overline{V},\underline{V}),\eta^{-N}\}\left(\sum_{r=t_1}^{t-1}\|w_r\|_{Q}\right)^2. \label{eq:proof_1}
		\end{align}
		Since $W(\hat{x}_{t-N},x_{t-N})\leq\Gamma(1,\hat{x}_{t-N},x_{t-N},\hat{z}_{t-N},z_{t-N})$
		by~\eqref{eq:candidate}, we can recursively apply \eqref{eq:proof_1} for $j$ times, which by exploiting the geometric series yields
		\begin{align}
			&\Gamma_1(t,\hat{x}_{t},x_{t},\hat{z}_{t},z_t) \nonumber\\ 
			&\leq \rho^jW(\hat{x}_{t_1},x_{t_1}) + \frac{2c_1'\overline{\lambda}(\overline{V},\underline{V})}{1-\rho} V(\bar{z}_{t_1},z_{t_1})^2\nonumber\\
			&\quad +  \frac{c_1'\max\{2\overline{\lambda}(\overline{V},\underline{V}),\eta^{-N}\}}{1-\rho} \left(\sum_{r={t_1}}^{t-1}\|w_{r}\|_{{Q}}\right)^2.\label{eq:proof_P1}
		\end{align}
		By~\eqref{eq:c_def}, we have that $2c_1'\overline{\lambda}(\overline{V},\underline{V})/(1-\rho)<c$, leading to \eqref{eq:proof_P1_res} with $c_{\mathrm{Q},1} = c_1'\max\{2\overline{\lambda}(\overline{V},\underline{V}),\eta^{-N}\}/(1-\rho)$, and we note that \eqref{eq:proof_P1_res} holds for any $j\in\mathbb{I}_{\geq0}$.
		
		It remains to establish the bound~\eqref{eq:proof_P2_res}. Assume that $k\in\mathbb{I}_{\geq1}$. Then, $t_1$ corresponds to the most recent horizon where $X_{t_1}\in\mathbb{E}_N$. We can invoke Lemma~\ref{lem:PE}, which yields
		\begin{align}
			&\Gamma(c,\hat{x}_{t_1},x_{t_1},\hat{z}_{t_1},z_{t_1})\nonumber\\
			&\ \leq \mu\Gamma(1,\bar{x}_{t_1-N},{x}_{t_1-N},\bar{z}_{t_1-N},z_{t_1-N})\nonumber\\
			&\ \quad  + 4c_{1}(c,N)\max\{1,\overline{\lambda}(\overline{V},S_o)\}\sum_{r={t_1-N}}^{t_1-1}\|w_{r}\|_{Q}^2. \label{eq:proof_P2_1}
		\end{align}
		We further have that
		\begin{align}
			&\Gamma(1,\bar{x}_{t_1-N},{x}_{t_1-N},\bar{z}_{t_1-N},z_{t_1-N}) \nonumber \\
			%&= U(\bar{x}_{t_1-N},{x}_{t_1-N}) + V(\bar{z}_{t_1-N},z_{t_1-N})\nonumber\\
			&\leq  \Gamma(1,\hat{x}_{t_1-N},{x}_{t_1-N},\hat{z}_{t_1-N},z_{t_1-N})  + V(\bar{z}_{t_1-N},z_{t_1-N})^2\nonumber\\
			&\leq  \Gamma(1,\hat{x}_{t_1-N},{x}_{t_1-N},\hat{z}_{t_1-N},z_{t_1-N})  \nonumber\\
			&\quad + 2V(\bar{z}_{t_2},z_{t_2})^2 + 2\left(\sum_{r=t_2}^{t_1-N-1}\|w_r\|_{Q_{\mathrm{v}}}\right)^2, \label{eq:proof_P2_2}
		\end{align}
		where in the latter inequality we have used similar arguments that were applied to derive~\eqref{eq:proof_V}.
		Adapting the arguments applied to derive~\eqref{eq:proof_P1}, we have that
		\begin{align}
			&\Gamma(1,\hat{x}_{t_1-N},x_{t_1-N},\hat{z}_{t_1-N},z_{t_1-N})\nonumber\\
			&\leq W(\hat{x}_{t_2},x_{t_2}) + \frac{2c_1'\overline{\lambda}(\overline{V},\underline{V})}{1-\rho}V(\bar{z}_{{t_2}},z_{t_2})^2\nonumber \\
			&\quad  +  \frac{c_1'\max\{2\overline{\lambda}(\overline{V},\underline{V}),\eta^{-N}\}}{1-\rho}\left(\sum_{r=t_2}^{t_1-N-1}\|w_{r}\|_{{Q}}\right)^2,\label{eq:proof_P2_3}
		\end{align}
		where we have used that $\rho^s\leq 1 $ for all $s\in\mathbb{I}_{\geq0}$.
		The combination of \eqref{eq:proof_P2_1}--\eqref{eq:proof_P2_3} (which also hold with $t_1$ and $t_2$ replaced by $t_m$ and $t_{m+1}$ for each $m\in \mathbb{I}_{[1,k]}$) and using the definition of $c$ from \eqref{eq:c_def} leads to \eqref{eq:proof_P2_res}, where
		\begin{align*}
			c_{\mathrm{Q},2} := \max\Bigg\{&\mu\left(\frac{c_1'\max\{2\overline{\lambda}(\overline{V},\underline{V}),\eta^{-N}\}}{1-\rho}+2\right),\\
			&\
			4c_{1}(c,N)\max\{1,\overline{\lambda}(\overline{V},S_o)\}
			\Bigg\},
		\end{align*}
		which finishes this proof. \qed
	\end{pf}

	\begin{thm}\label{thm:nonPE}
		Let the system~\eqref{eq:sys} be i-IOSS according to Definition~\ref{ass:W} and Assumption~\ref{ass:V} be satisfied. Consider the MHE scheme~\eqref{eq:MHE} with cost function \eqref{eq:MHE_objective}.
		Suppose that $\eta$ and $N$ satisfy~\eqref{eq:contraction}.
		Then, the estimates~\eqref{eq:MHE_estimates} satisfy
		\begin{align}
			&C_0\|\hat{x}_{t}-x_{t}\|_{\underline{W}} + C_0\|\hat{z}_{t}-z_t\|_{\underline{V}}\nonumber\\
			&\leq \sqrt{\mu}^{k}\Big(\sqrt{C_1}\sqrt{\tilde{\eta}}^l\|\hat{x}_{0}-x_{0}\|_{\overline{W}} + \sqrt{C_2}\sqrt{\eta}^l \|\hat{z}_{0}-z_0\|_{\overline{V}}\Big)\nonumber\\
			&\quad + \sum_{r={t_1}}^{t-1}\|w_{r}\|_{{Q_3}} +\sum_{m=1}^k \sqrt{\mu}^{m-1}\hspace{-1ex}\sum_{r={t_{m+1}}}^{t_m-1} \|w_{r}\|_{Q_3} \nonumber\\
			&\quad + \sqrt{\mu}^{k}\sum_{r={0}}^{l-1}\|w_{r}\|_{{Q_3}}\label{eq:thm_nonPE}
		\end{align}
		for all $t\in\mathbb{I}_{\geq0}$ and all $\hat{x}_0,x_0\in\mathbb{X}$, all $\hat{z}_0,z_0\in\mathbb{Z}$, and every sequence $\{u_r\}_{r=0}^{\infty}\in\mathbb{U}^\infty$ and $\{w_r\}_{r=0}^{\infty}\in\mathbb{W}^\infty$, where $\tilde{\eta}:= \max\{\eta_{\mathrm{w}},\eta_{\mathrm{o}}\}$, and $C_0 := \sqrt{2}/2$,
		\begin{align}
			C_1 &:= 2\eta^{-N}c_{1}(1,N)(2+ \overline{\lambda}(P_\mathrm{o},\overline{U})),\label{eq:C1_def}\\
			C_2 &:= (4c_1(1,N)+2c)\eta^{-N}, \label{eq:C2_def}\\
			Q_3&:=\max\{c_{\mathrm{Q},1},c_{\mathrm{Q},2},(4c_1(1,N)\eta^{-N}+2c)\}Q.\label{eq:Q_def}
		\end{align}
	\end{thm}

	Before proving Theorem~\ref{thm:nonPE}, we want to highlight the key properties of the resulting bound on the estimation errors.
	
	\begin{itemize}
		\item The bound~\eqref{eq:thm_nonPE} is valid independent of the observability of the parameter $z$, and it improves the more often $z$ is observable during operation.
		\item If $k\rightarrow\infty$ for $t\rightarrow\infty$, then $e_t$ \eqref{eq:MHE_error} converges to a region around the origin defined by the true disturbances. If additionally $w\rightarrow0$ for $t\rightarrow\infty$,~then~$e_t\rightarrow0$.
		\item If $t-t_1$ and $t_{m+1}-t_m$ can be uniformly bounded for all times, then the estimation error converges exponentially, which follows by using similar arguments as in~\cite[Cor.~10]{Schiller2023d}.
		\item In contrast to our results for constant parameters \citep{Schiller2023d}, the bound~\eqref{eq:thm_nonPE} involves the (non-discounted) sum of disturbances $w$ over $[t_1,t]$, which can be interpreted as the energy of the corresponding disturbance subsequence $\{w_r\}_{r=t_1}^t$.
		This is due to the fact that $z$ is unobservable in this interval, where we use the nominal dynamics of $\eqref{eq:sys_z}$ to predict the evolution of $z$ and invoke Assumption~\ref{ass:V} to bound the current parameter estimation error $\|e_{z,t}\|$.
		Hence, no qualitatively better error bound can be expected without further assumptions on the dynamics~\eqref{eq:sys_z}, the disturbances~$w$, or observability with respect to $z$.
	\end{itemize}

	\begin{pf*}{Proof of Theorem~\ref{thm:nonPE}.}	
		The claim follows by combining the bounds from Lemma~\ref{lem:bounds} and invoking Assumption~\ref{ass:V} for $l\in\mathbb{I}_{[0,N-1]}$.
		First, suppose that $k\in\mathbb{I}_{\geq1}$ and note that $\bar{z}_{t_m}=\hat{z}_{t_m}$ for all $m\in\mathbb{I}_{[1,k]}$.
		%\JS{$\tau=\min\{1,k\}t_1$ either satisfies $\tau=t_1$ if $k\in\mathbb{I}_{\geq0}$ or $\tau=0$ if $k=0$}
		Hence, the recursive application of \eqref{eq:proof_P2_res} and the fact that $t_{k+1} = l$ yield
		\begin{align}
			&\Gamma(c,\hat{x}_{t_1},x_{t_1},\bar{z}_{t_1},z_{t_1})\nonumber\\ 
			&\leq \mu^{k} \Gamma(c,\hat{x}_{l},x_{l},\bar{z}_{l},z_l) +\sum_{m=1}^k \mu^{m-1} \left(c_{Q,2}\sum_{r={t_{m+1}}}^{t_m-1}\hspace{-1ex} \|w_r\|_{Q}\right)^2, \label{eq:proof_P3_res}
		\end{align}
		which is valid for all $k\geq0$ (in case $k=0$, we have $t_1=l$).
		For $l\in\mathbb{I}_{[0,N-1]}$, it follows that
		\begin{align}
			\Gamma(c,\hat{x}_l,x_l,\bar{z}_l,z_l) &= W(\hat{x}_l,x_l) + cV(\bar{z}_l,z_l)^2\nonumber\\
			&\leq \Gamma(1,\hat{x}_l,x_l,\hat{z}_l,z_l) + cV(\bar{z}_l,z_l)^2. \label{eq:proof_P4_1}
		\end{align}
		By using Lemma~\ref{lem:nonPE} with $\bar{x}_0=\hat{x}_0$ and $\bar{z}_0=\hat{z}_0$, we obtain
		\begin{align}
			&\Gamma(1,\hat{x}_l,x_l,\hat{z}_l,z_l)\nonumber\\
			&\leq c_{1}(1,N)\eta^{-N}(2\eta_\mathrm{w}^{l}+2\gamma(l))\|\hat{x}_{0}-x_{0}\|_{\overline{W}}^2\nonumber\\
			&\quad + c_1'\eta^{-N}\eta^l\|\hat{z}_{0}-z_0\|^2_{\overline{V}}+ c_1'\eta^{-N}\sum_{r={0}}^{l-1}\|w_{r}\|_{Q}^2, \label{eq:proof_P4_2}
		\end{align}
		where we have used the definition $c_1'=4c_1(1,N)$ together with the facts that $l<N$ and $c_1(1,s)< c_1(1,N)$ for all $s\in[0,N-1]$.
		By Assumption~\ref{ass:V}, the facts that $\bar{z}_l=g^{(l)}(\bar{z}_0)$ and $\bar{z}_0=\hat{z}_0$, and Jensen's inequality, it further follows that
		\begin{equation}
			V(\bar{z}_l,z_l)^2 \leq 2V(\bar{z}_0,z_0)^2 + 2\left(\sum_{r=0}^{l-1}\|w_r\|_{Q_\mathrm{v}}\right)^2. \label{eq:proof_P4_3}
		\end{equation}
		From \eqref{eq:proof_P4_1}--\eqref{eq:proof_P4_3} and the definitions of $\gamma$, $C_1$, and $C_2$ from~\eqref{eq:gamma}, \eqref{eq:C1_def}, and~\eqref{eq:C2_def}, respectively, we can infer that
		\begin{align}
			\Gamma(c,\hat{x}_{l},x_{l},\bar{z}_{l},z_l) \leq&\ C_1\tilde{\eta}^l\|\hat{x}_{0}-x_{0}\|_{\overline{W}}^2 + C_2\eta^l \|\hat{z}_{0}-z_0\|^2_{\overline{V}}\nonumber\\
			& + (c_1'\eta^{-N}+2c)\left(\sum_{r={0}}^{l-1}\|w_{r}\|_{{Q}}\right)^2. \label{eq:proof_P4_res}
		\end{align}
		Combining \eqref{eq:proof_P1_res}, \eqref{eq:proof_P3_res}, and \eqref{eq:proof_P4_res}, using the definition of $Q_3$ from~\eqref{eq:Q_def}, applying the square root (which is concave and sub-additive on $\mathbb{R}_{\geq0}$), and Jensen's inequality finally leads to~\eqref{eq:thm_nonPE}.
		Since~\eqref{eq:thm_nonPE} holds for any $l\in\mathbb{I}_{[0,N-1]}$ and $k\in\mathbb{I}_{\geq0}$, it holds for all $t\in\mathbb{I}_{\geq0}$, which finishes this proof. \qed
	\end{pf*}

	We conclude the paper with a simulation example in the next section.

	\section{Numerical example}
	We consider the following system
	\begin{align*}
		x_1^+ &= x_1 + t_{\Delta}b_1(x_2-a_1x_1-a_2x_1^2-zx_1^3) + w_1,\\
		x_2^+ &= x_2 + t_{\Delta}(x_1-x_2+x_3) + w_2,\\
		x_3^+ &= x_3-t_{\Delta}b_2x_2 + w_3,\\
		z^+ &= z + w_4\\
		y &= x_1 + w_5.
	\end{align*}
	This corresponds to the Euler-discretized modiﬁed Chua's circuit system from~\cite{Yang2015} using the step size $t_{\Delta} = 0.01$, where we consider additional process disturbances, noisy output measurements, and assume that the parameter $z_t$ is time-variant and subject to an unknown disturbance input $w_4$. Note that from the system description it is immediately apparent that the observability of $z$ depends on the magnitude of $x_1$.
	The parameters are $b_1=12.8$, $b_2 = 19.1$, $a_1=0.6$, $a_2=-1.1$.
	The disturbances $w_i$, $i=1,2,3,5$ are uniformly distributed with $|w_i|\leq 10^{-3}, i=1,2,3$ for the process disturbance, and $|w_5|\leq 5\cdot10^{-2}$ for the measurement noise. The disturbance $w_4$ consists of two superimposed square waves such that $w_4\in \{-10^{-4},0,10^{-4}\}$, compare the red curve in Figure~\ref{fig:2} below.
	We consider the initial conditions $x_0=[2,0,-1]^\top$ and $z_0=0.45$ and assume that $x_t$ and $z_t$ evolve in the (known) sets $\mathbb{X} = [-1,3]\times[-1,1]\times[-3,3]$ and $\mathbb{Z}=[0.2,0.8]$.
	The objective is to compute the state and parameter estimates $\hat{x}_t$ and $\hat{z}_t$ by applying the MHE scheme proposed in Section~\ref{sec:MHE} using $\hat{x}_0=[-1,0.1,2]^\top$~and~$\hat{z}_0=0.6$.
	
	Assumption~\ref{ass:V} is satisfied with $V(z,\tilde{x})=\|z-\tilde{z}\|$, $\mathrm{Q}_v = B_{\mathrm{z}}^\top B_{\mathrm{z}}$, and $B_{\mathrm{z}} = [0, 0, 0, 1, 0]$.
	We compute a quadratic \mbox{i-IOSS} Lyapunov function $W(x,\tilde{x}) = \|x-\tilde{x}\|_{P_\mathrm{w}}^2$, $P_{\mathrm{w}}\succ0$ using the method from~\cite{Schiller2023c}.
	To monitor the observability of trajectory pairs, we construct at each time $t\in\mathbb{I}_{\geq0}$ a matrix $\mathcal{O}_{N_t} = \sum_{j=0}^{N_t-1}\mu^{N_t-j+1} Y_{j}^\top C^\top CY_j$ for some suitable $\mu\in(0,1)$ and $C = [1, 0, 0]$, where $Y_{j+1} =\Phi Y_j + {\partial f / \partial z}(\hat{x}^*_{t-N+j|t})$ for $j\in\mathbb{I}_{[0,N_t-1]}$, $Y_0  = 0_{n_z\times n_z}$, and $\Phi$ is a constant Schur-stable matrix.
	We consider a trajectory pair $X_t$ to be observable if $\alpha_t = \underline{\lambda}(\mathcal{O}_{N_t})\geq \alpha= 5\cdot10^{-4}$.
	For further details on the construction of $\mathcal{O}_T$, see \citep{Schiller2023d}.
	
	Tuning the cost function~\eqref{eq:MHE_objective} based on the parameters of $W$, $V$, and the set $\mathbb{E}_{T}$ yields a minimum required horizon length $N_{\min}\approx 300$. This is rather conservative due to conservative steps in the proofs (in particular, Lemmas~\ref{lem:nonPE} and~\ref{lem:PE}), and good performance is obtained using a smaller horizon length.
	To illustrate the potential of the proposed MHE scheme, we choose $N=200$ and $\eta = 0.9$, $\gamma(s)=\eta^s$, $\overline{W}=I_{2}$, $\overline{V}=1$, ${Q}=10^7I_{4}$, ${R}=10^3$, although these invalidate the theoretical guarantees established in Section~\ref{sec:MHE_analysis}. Furthermore, we update $\hat{z}_t$ only if $X_t\in\mathbb{E}_{N_t}$, compare \cite[Rem.~9]{Schiller2023d}.

		\begin{figure}
		\flushright
		\includegraphics{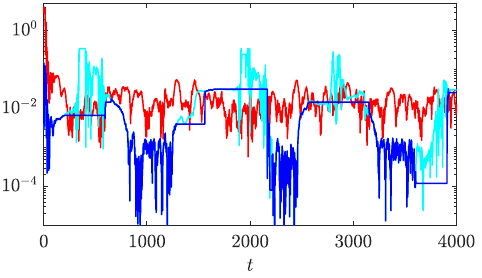}
		\caption{Estimation errors $\|e_{\mathrm{x},t}\|$ (red) and $\|e_{\mathrm{z},t}\|$ for the proposed MHE scheme (blue) compared to MHE without observability monitoring (cyan).}
		\label{fig:1}
	\end{figure}
	\begin{figure}
		\flushright
		\includegraphics{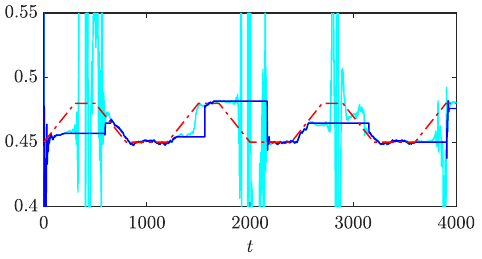}
		\caption{Estimates $\hat{z}_t$ for the proposed MHE scheme (blue) compared to MHE without observability monitoring (cyan) and the true parameter $z_t$ (red).}
		\label{fig:2}
	\end{figure}
	\begin{figure}
		\flushright
		\includegraphics{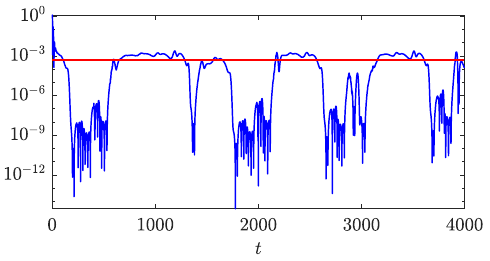}
		\caption{Observability level $\alpha_t$. Values of $\alpha_t$ above the red line indicate an observable trajectory pair at time $t$.}
		\label{fig:3}
	\end{figure}
	The simulations are carried out in Matlab over $t_\mathrm{sim}=4000$ time steps.
	The estimation error of the proposed scheme is shown in Figure~\ref{fig:1}, which shows fast convergence of the estimation errors. This is also evident in Figure~\ref{fig:2}, which illustrates the respective estimated parameter $\hat{z}_t$ over time.
	In phases without sufficiently high observability levels ($\alpha_t\ll\alpha$, cf. Figure~\ref{fig:3}), the proposed MHE scheme is not able to track the true parameter $z_t$ (which is clear due to the lack of information). However, it still provides estimates with bounded errors that are much smaller compared to using MHE without observability monitoring (see the cyan-colored curve in Figures~\ref{fig:1} and \ref{fig:2}).
	
	This simulation example illustrates the efficiency of the proposed MHE scheme from Section~\ref{sec:MHE}. When $z$ is observable from data, it is able to robustly track the true parameter. If it is unobservable, it reacts appropriately to prevent the estimation error from deteriorating arbitrarily.

	\begingroup
	\small

	\endgroup
		
	\appendix
	\section{Proof of Lemma~\ref{lem:nonPE}}\label{sec:App1}
		\begin{pf}
		From~\eqref{eq:candidate} and the optimal estimates~\eqref{eq:MHE_estimates}, we obtain
		\begin{equation}
			\Gamma(1,\hat{x}_t,x_t,\hat{z}_t,z_t)=  W(\hat{x}_{t|t}^*,x_{t}) + V(\hat{z}^*_{t|t},z_t)^2. \label{eq:proof_start}
		\end{equation}
		Satisfaction of the MHE constraints~\eqref{eq:MHE_x}--\eqref{eq:MHE_con} implies that the optimal trajectory satisfies the dynamics~\eqref{eq:sys} and $(\hat{x}_{j|t}^*,\hat{z}_{j|t}^*,u_j,{w}_{j|t}^*)\in\mathbb{D}$ for all $j\in\mathbb{I}_{[t-N_t,t-1]}$.
		Invoking the property~\eqref{eq:W_dissip} yields
		\begin{align*}
			&W(\hat{x}_{t|t}^*,x_{t}) + V(\hat{z}_{t|t}^*,z_{t})^2\nonumber\\
			&\leq  \eta_\mathrm{w}^{N_t}
			W(\hat{x}_{t-N_t|t}^*,x_{t-N_t}) \nonumber\\
			&\quad+ \sum_{j={1}}^{N_t}\eta_{\mathrm{w}}^{j-1}(\|\hat{w}^*_{t-j|t}-w_{t-1}\|^2_{Q_\mathrm{w}} + \|\hat{y}^*_{t-j|t}-y_{t-j}\|_{{R}_\mathrm{w}}^2)\nonumber\\
			&\quad + \sum_{j={1}}^{N_t}\eta_{\mathrm{w}}^{j-1}\|\hat{z}^*_{t-j|t}-z_{t-j}\|^2_{S_\mathrm{w}} + V(\hat{z}^*_{t|t},z_{t})^2.
		\end{align*}
		Exploiting the bound~\eqref{eq:V_bounds}, we obtain
		\begin{align*}
			& V(\hat{z}_{t|t}^*,z_{t})^2 + \sum_{j={1}}^{N_t}\eta_{\mathrm{w}}^{j-1}\|\hat{z}^*_{t-j|t}-z_{t-j}\|^2_{S_\mathrm{w}}\\
			& \leq  \max\{1,\overline{\lambda}(S_{\mathrm{w}},\underline{V})\eta_{\mathrm{w}}^{-1}\} \sum_{j={0}}^{N_t}\eta_{\mathrm{w}}^{j}V(\hat{z}^*_{t-j|t},z_{t-j})^2.
		\end{align*}
		From~\eqref{eq:V_dissip}, we further have that for each $j\in\mathbb{I}_{[0,N_t]}$,
		\begin{align*}
			&V(\hat{z}^*_{t-j|t},z_{t-j})^2\\
			%& \leq \Big( V(\hat{z}^*_{t-N_t|t},z_{t-N_t}) + \sum_{i=j+1}^{N_t}\|\hat{w}^*_{t-i|t}-w_{t-i}\|_{Q_\mathrm{v}}\Big)^2\\
			%& \leq \Big( V(\hat{z}^*_{t-N_t|t},z_{t-N_t}) + \sum_{i=1}^{N_t}\|\hat{w}^*_{t-i|t}-w_{t-i}\|_{Q_\mathrm{v}}\Big)^2\\
			& \leq (N_t+1) \Big(V(\hat{z}^*_{t-N_t|t},z_{t-N_t})^2 + \sum_{i=1}^{N_t}\|\hat{w}^*_{t-i|t}-w_{t-i}\|_{Q_\mathrm{v}}^2\Big)
		\end{align*}
		using Jensen's inequality. Combining the previous inequalities and using the geometric series, we obtain that
		\begin{align*}
			&W(\hat{x}_{t|t}^*,x_{t}) + V(\hat{z}_{t|t}^*,z_{t})^2\nonumber\\
			&\leq  \eta_\mathrm{w}^{N_t}
			W(\hat{x}_{t-N_t|t}^*,x_{t-N_t}) \nonumber\\
			&\ \, + \sum_{j={1}}^{N_t}\eta_{\mathrm{w}}^{j-1}(\|\hat{w}^*_{t-j|t}-w_{t-1}\|^2_{Q_\mathrm{w}} + \|\hat{y}^*_{t-j|t}-y_{t-j}\|_{{R}_\mathrm{w}}^2)\nonumber\\
			&\ \, + c_1(1,N_t) \Big(V(\hat{z}^*_{t-N_t|t},z_{t-N_t})^2 + \sum_{i=1}^{N_t}\|\hat{w}^*_{t-i|t}-w_{t-i}\|_{Q_\mathrm{v}}^2\Big)
		\end{align*}
		with $c_1(r,s)$ from~\eqref{eq:c1_s}.
		Using the fact that $c_1(1,s)\geq1$ for all $s\geq0$, it follows that
		\begin{align}
			&W(\hat{x}_{t|t}^*,x_{t}) + V(\hat{z}_{t|t}^*,z_{t})^2\nonumber\\
			&\leq  c_1(1,N_t) \Big(\eta_\mathrm{w}^{N_t} 	
			W(\hat{x}_{t-N_t|t}^*,x_{t-N_t}) + V(\hat{z}^*_{t-N_t|t},z_{t-N_t})^2\nonumber\\
			&\quad + \sum_{j=1}^{N_t}\|\hat{w}^*_{t-j|t}-w_{t-j}\|^2_{Q_\mathrm{w} + Q_\mathrm{v}} + \|\hat{y}^*_{t-j|t}-y_{t-j}\|_{{R}_\mathrm{w}}^2\Big). \label{eq:proof_start_2}
		\end{align}

		Using the bounds~\eqref{eq:W_bounds} and~\eqref{eq:V_bounds} together with Jensen's inequality, we obtain that
		\begin{align}
			&W(\hat{x}_{t-N_t|t}^*,x_{t-N_t}) \leq \|\hat{x}^*_{t-N_t|t}-x_{t-N_t}\|^2_{\overline{W}}\nonumber\\
			&\ \leq 2\|\hat{x}^*_{t-N_t|t}-\bar{x}_{t-N_t}\|^2_{\overline{W}} + 2\|\bar{x}_{t-N_t}-x_{t-N_t}\|^2_{\overline{W}}, \label{eq:proof_triangle_x}\\
			&V(\hat{z}_{t-N_t|t}^*,z_{t-N_t})^2 \leq \|\hat{z}^*_{t-N_t|t}-z_{t-N_t}\|^2_{\overline{V}}\nonumber\\
			&\ \leq 2\|\hat{z}^*_{t-N_t|t}-\bar{z}_{t-N_t}\|^2_{\overline{V}} + 2\|\bar{z}_{t-N_t}-z_{t-N_t}\|^2_{\overline{V}},
			\label{eq:proof_triangle_z}\\
			&\|\hat{w}^*_{t-j|t}-w_{t-j}\|^2_{\tilde{Q}} \leq 2\|\hat{w}^*_{t-j|t}\|^2_{\tilde{Q}} + 2\|w_{t-j}\|^2_{\tilde{Q}} \label{eq:proof_triangle_w}
		\end{align}
		for all $j\in\mathbb{I}_{[1,N_t]}$ with $\tilde{Q}=Q_\mathrm{w} + Q_\mathrm{v}$. Using~\eqref{eq:proof_start_2}--\eqref{eq:proof_triangle_w} in \eqref{eq:proof_start}, the facts that $\eta^{N_t-N} \geq 1$ and $c_1(1,N_t)> 1$, and the definition of the cost function~\eqref{eq:MHE_objective} lets us infer that
		\begin{align}
		&\Gamma(1,\hat{x}_t,x_t,\hat{z}_t,z_t)\nonumber\\
		&\leq  \eta^{-N}c_1(1,N_1)\Big(2\eta_\mathrm{w}^{N_t} 	
		\|\bar{x}_{t-N_t}-{x}_{t-N_t}\|_{\overline{W}}\nonumber\\[-0.3ex]
		&\quad + 2\eta^{N_t}\|\bar{z}_{t-N_t}-{z}_{t-N_t}\|^2_{\overline{V}} + \sum_{j=1}^{N_t}2\|w_{t-j}\|^2_{Q}\nonumber\\
		&\quad + J(\hat{x}_{t-N_t|t}^*,\hat{z}_{t-N_t|t}^*,\hat{w}_{\cdot|t}^*)\Big). \label{eq:proof_cost}
		\end{align}
		By optimality, it follows that
		\begin{align}
		&J_t(\hat{x}_{t-N_t|t}^*,\hat{z}_{t-N_t|t}^*,\hat{w}_{\cdot|t}^*) \leq J_t(x_{t-N_t},z_{t-N_t},w_{\cdot|t})\nonumber\\
		&\ \leq 2\gamma(N_t)\|x_{t-N_t}-\bar{x}_{t-N_t}\|_{\overline{W}}^2 +  2{\eta}^{N_t}\|z_{t-N_t}-\bar{z}_{t-N_t}\|_{\overline{V}}^2\nonumber\\[-0.3ex]
		&\ \quad + 2\sum_{j=1}^{N_t}\|{w}_{t-j}\|_{Q}^2. \label{eq:proof_optimality}
		\end{align}
		Combining \eqref{eq:proof_cost} and \eqref{eq:proof_optimality} yields \eqref{eq:lem_nonPE}, which finishes this proof. \qed
	\end{pf}

	\section{Proof of Lemma~\ref{lem:PE}}\label{sec:App2}
		\begin{pf}
		We start by using the same arguments as in the beginning of the proof of Lemma~\ref{lem:nonPE}, which leads to
		\begin{align*}
		&\Gamma(c,\hat{x}_{t},x_t,\hat{z}_{t},z_t) =  W(\hat{x}_{t|t}^*,x_t) + cV(\hat{z}_{t|t}^*,z_t)^2\\
		&\leq  c_1(c,N)\Big(\eta_\mathrm{w}^{N} 	
		W(\hat{x}_{t-N|t}^*,x_{t-N}) + V(\hat{z}^*_{t-N|t},z_{t-N})^2\\
		&\quad + \sum_{j=1}^{N}\|\hat{w}^*_{t-j|t}-w_{t-j}\|^2_{Q_\mathrm{w} + Q_\mathrm{v}} + \|\hat{y}^*_{t-j|t}-y_{t-j}\|_{R_\mathrm{w}}^2\Big).
		\end{align*}
		Using that $V(\hat{z}^*_{t-N|t},z_{t-N})^2\leq \overline{\lambda}(\overline{V},S_o)\|\hat{z}^*_{t-N|t}-z_{t-N}\|^2_{S_\mathrm{o}}$
%		\begin{align*}
%		& \|\hat{z}_{t-N|t}^*-z_{t-N}\|^2_{S_{\mathrm{o}}} \leq\eta_\mathrm{o}^N\|\hat{x}_{t-N|t}^*-{x}_{t-N}\|^2_{P_\mathrm{o}} \\
%		&+ \sum_{j={1}}^N(\|\hat{w}_{t-j|t}^*-{w}_{t-j}\|^2_{Q_{\mathrm{o}}} + \|\hat{y}_{t-j|t}^*-{y}_{t-j}\|^2_{R_\mathrm{o}})
%		\end{align*}
		and invoking that~$X_t\in\mathbb{E}_N$, we can write that
		\begin{align*}
		&\Gamma(c,\hat{x}_{t},x_t,\hat{z}_{t},z_t)\\
		&\ \leq  c_1(c,N)\max\{1,\overline{\lambda}(\overline{V},S_o)\}\Big(\eta_\mathrm{w}^{N} 	
		W(\hat{x}_{t-N|t}^*,x_{t-N}) \\
		&\ \quad +
		\eta_\mathrm{o}^N\|\hat{x}_{t-N|t}^*-{x}_{t-N}\|^2_{P_o} \\[-0.3ex]
		&\ \quad + \sum_{j={1}}^N\|\hat{w}_{t-j|t}^*-{w}_{t-j}\|^2_{Q} + \|\hat{y}_{t-j|t}^*-{y}_{t-j}\|^2_{R},
		\end{align*}
		where we recall that ${Q} = Q_\mathrm{w} + Q_\mathrm{v} + Q_\mathrm{o}$ ${R} = R_\mathrm{w}+R_\mathrm{o}$.
		Using the definition of $\gamma$ from~\eqref{eq:gamma}, the bounds from \eqref{eq:proof_triangle_x} and \eqref{eq:proof_triangle_w}, and the cost function~\eqref{eq:MHE_objective}, we have that
		\begin{align*}
		&\Gamma(c,\hat{x}_{t},x_t,\hat{z}_{t},z_t)\\
		&\leq  c_1(c,N)\max\{1,\overline{\lambda}(\overline{V},S_o)\}\Big(
		2\gamma(N)\|\bar{x}_{t-N|t}-{x}_{t-N}\|^2_{\overline{W}}\\[-0.3ex]
		&\quad + 2\sum_{j=1}^{N}\|w_{t-j}\|^2_{Q}  + J(\hat{x}_{t-N|t}^*,\hat{z}_{t-N|t}^*,\hat{w}_{\cdot|t}^*)\Big).
		\end{align*}
		Exploiting optimality as in~\eqref{eq:proof_optimality} and using the definition of $\mu$ from~\eqref{eq:mu} yields \eqref{eq:lem_PE}, which finishes this proof. \qed
%		\begin{align*}
%		\Gamma(c,\hat{x}_{t},x_t,\hat{z}_{t},z_t)\leq &\ c_1(c,N)\max\{1,\overline{\lambda}(\overline{V},S_o)\}\\
%		&\cdot\Big(
%		\gamma(N)4\overline{\lambda}(\overline{W},\underline{W})W(\bar{x}_{t-N|t},{x}_{t-N})\\
%		&\quad + 2\eta^N\overline{\lambda}(\overline{V},\underline{V})V(\bar{z}_{t-N|t},{z}_{t-N})^2\\
%		&\quad +  4\sum_{j=1}^{N}\|w_{t-j}\|^2_{Q}\Big).
%		\end{align*}
%		Using the definition of $\mu$ from~\eqref{eq:mu} yields \eqref{eq:lem_PE}, which finishes this proof. \qed
	\end{pf}	
		

\begin{thebibliography}{25}
\providecommand{\natexlab}[1]{#1}
\providecommand{\url}[1]{\texttt{#1}}
\providecommand{\urlprefix}{URL }
\expandafter\ifx\csname urlstyle\endcsname\relax
  \providecommand{\doi}[1]{doi:\discretionary{}{}{}#1}\else
  \providecommand{\doi}{doi:\discretionary{}{}{}\begingroup
  \urlstyle{rm}\Url}\fi

\bibitem[{Alessandri et~al.(2012)Alessandri, Baglietto, and
  Battistelli}]{Alessandri2012}
Alessandri, A., Baglietto, M., and Battistelli, G. (2012).
\newblock Min-max moving-horizon estimation for uncertain discrete-time linear
  systems.
\newblock \emph{{SIAM} J. Control Optim.}, 50(3), 1439--1465.

\bibitem[{Alessandri et~al.(2008)Alessandri, Baglietto, and
  Battistelli}]{Alessandri2008}
Alessandri, A., Baglietto, M., and Battistelli, G. (2008).
\newblock Moving-horizon state estimation for nonlinear discrete-time systems:
  New stability results and approximation schemes.
\newblock \emph{Automatica}, 44(7), 1753--1765.

\bibitem[{Allan and Rawlings(2021)}]{Allan2021a}
Allan, D.A. and Rawlings, J.B. (2021).
\newblock Robust stability of full information estimation.
\newblock \emph{{SIAM} J. Control Optim.}, 59(5), 3472--3497.

\bibitem[{Allan et~al.(2021)Allan, Rawlings, and Teel}]{Allan2021}
Allan, D.A., Rawlings, J.B., and Teel, A.R. (2021).
\newblock Nonlinear detectability and incremental input/output-to-state
  stability.
\newblock \emph{{SIAM} J. Control Optim.}, 59(4), 3017--3039.

\bibitem[{Angeli et~al.(2000{\natexlab{a}})Angeli, Sontag, and
  Wang}]{Angeli2000}
Angeli, D., Sontag, E.D., and Wang, Y. (2000{\natexlab{a}}).
\newblock A characterization of integral input-to-state stability.
\newblock \emph{{IEEE} Trans. Autom. Control}, 45(6), 1082--1097.

\bibitem[{Angeli et~al.(2000{\natexlab{b}})Angeli, Sontag, and
  Wang}]{Angeli2000a}
Angeli, D., Sontag, E.D., and Wang, Y. (2000{\natexlab{b}}).
\newblock Further equivalences and semiglobal versions of integral input to
  state stability.
\newblock \emph{Dyn. Control}, 10(2), 127--149.

\bibitem[{Arezki et~al.(2023)Arezki, Zemouche, Alessandri, and
  Bagnerini}]{Arezki2023}
Arezki, H., Zemouche, A., Alessandri, A., and Bagnerini, P. (2023).
\newblock {LMI} design procedure for incremental input/output-to-state
  stability in nonlinear systems.
\newblock \emph{{IEEE} Control Syst. Lett}, 7, 3403--3408.

\bibitem[{Bastin and Gevers(1988)}]{Bastin1988}
Bastin, G. and Gevers, M. (1988).
\newblock Stable adaptive observers for nonlinear time-varying systems.
\newblock \emph{{IEEE} Trans. Autom. Control}, 33(7), 650--658.

\bibitem[{Farza et~al.(2009)Farza, M'Saad, Maatoug, and Kamoun}]{Farza2009}
Farza, M., M'Saad, M., Maatoug, T., and Kamoun, M. (2009).
\newblock Adaptive observers for nonlinearly parameterized class of nonlinear
  systems.
\newblock \emph{Automatica}, 45(10), 2292--2299.

\bibitem[{Flayac and Shames(2023)}]{Flayac2023}
Flayac, E. and Shames, I. (2023).
\newblock Nonuniform observability for moving horizon estimation and stability
  with respect to additive perturbation.
\newblock \emph{SIAM J. Control Optim.}, 61(5), 3018--3050.

\bibitem[{Haimovich and Mancilla-Aguilar(2020)}]{Haimovich2020}
Haimovich, H. and Mancilla-Aguilar, J.L. (2020).
\newblock Strong {ISS} implies strong i{ISS} for time-varying impulsive
  systems.
\newblock \emph{Automatica}, 122, 109224.

\bibitem[{Hu(2024)}]{Hu2023}
Hu, W. (2024).
\newblock Generic stability implication from full information estimation to
  moving-horizon estimation.
\newblock \emph{{IEEE} Trans. Autom. Control}, 69(2), 1164--1170.

\bibitem[{Ioannou and Sun(2012)}]{Ioannou2012}
Ioannou, P. and Sun, J. (2012).
\newblock \emph{Robust Adaptive Control}.
\newblock Dover Publications, Inc., Mineola, NY, USA.

\bibitem[{Knüfer and Müller(2023)}]{Knuefer2023}
Knüfer, S. and Müller, M.A. (2023).
\newblock Nonlinear full information and moving horizon estimation: Robust
  global asymptotic stability.
\newblock \emph{Automatica}, 150, 110603.

\bibitem[{Liu et~al.(2022)Liu, Russo, Liberzon, and Cavallo}]{Liu2022a}
Liu, S., Russo, A., Liberzon, D., and Cavallo, A. (2022).
\newblock Integral-input-to-state stability of switched nonlinear systems under
  slow switching.
\newblock \emph{IEEE Trans. Autom. Control}, 67(11), 5841--5855.

\bibitem[{Marino et~al.(2001)Marino, Santosuosso, and Tomei}]{Marino2001}
Marino, R., Santosuosso, G., and Tomei, P. (2001).
\newblock Robust adaptive observers for nonlinear systems with bounded
  disturbances.
\newblock \emph{{IEEE} Trans. Autom. Control}, 45(6), 967--972.

\bibitem[{Muntwiler et~al.(2023)Muntwiler, Köhler, and
  Zeilinger}]{Muntwiler2023}
Muntwiler, S., Köhler, J., and Zeilinger, M.N. (2023).
\newblock {MHE} under parametric uncertainty -- robust state estimation without
  informative data.
\newblock \emph{arXiv:2312.14049}.

\bibitem[{Rao et~al.(2003)Rao, Rawlings, and Mayne}]{Rao2003}
Rao, C., Rawlings, J., and Mayne, D. (2003).
\newblock Constrained state estimation for nonlinear discrete-time systems:
  stability and moving horizon approximations.
\newblock \emph{{IEEE} Trans. Autom. Control}, 48(2), 246--258.

\bibitem[{Rawlings et~al.(2020)Rawlings, Mayne, and Diehl}]{Rawlings2017}
Rawlings, J.B., Mayne, D.Q., and Diehl, M.M. (2020).
\newblock \emph{Model Predictive Control: Theory, Computation, and Design}.
\newblock Nob Hill Publish., LLC, Santa Barbara, CA, USA, 2nd edition.
\newblock 3rd printing.

\bibitem[{Schiller et~al.(2023)Schiller, Muntwiler, Köhler, Zeilinger, and
  Müller}]{Schiller2023c}
Schiller, J.D., Muntwiler, S., Köhler, J., Zeilinger, M.N., and Müller, M.A.
  (2023).
\newblock A {L}yapunov function for robust stability of moving horizon
  estimation.
\newblock \emph{{IEEE} Trans. Autom. Control}, 68(12), 7466--7481.

\bibitem[{Schiller and Müller(2023)}]{Schiller2023d}
Schiller, J.D. and Müller, M.A. (2023).
\newblock Nonlinear moving horizon estimation for robust state and parameter
  estimation.
\newblock \emph{arXiv:2312.13175}.

\bibitem[{Sui and Johansen(2011)}]{Sui2011}
Sui, D. and Johansen, T.A. (2011).
\newblock Moving horizon observer with regularisation for detectable systems
  without persistence of excitation.
\newblock \emph{Int. J. Control}, 84(6), 1041--1054.

\bibitem[{{\c{T}}iclea and Besan{\c{c}}on(2016)}]{Ticlea2016}
{\c{T}}iclea, A. and Besan{\c{c}}on, G. (2016).
\newblock Adaptive observer design for discrete time {LTV} systems.
\newblock \emph{Int. J. Control}, 89(12), 2385--2395.

\bibitem[{Wynn et~al.(2014)Wynn, Vukov, and Diehl}]{Wynn2014}
Wynn, A., Vukov, M., and Diehl, M. (2014).
\newblock Convergence guarantees for moving horizon estimation based on the
  real-time iteration scheme.
\newblock \emph{{IEEE} Trans. Autom. Control}, 59(8), 2215--2221.

\bibitem[{Yang and Zhao(2015)}]{Yang2015}
Yang, J. and Zhao, L. (2015).
\newblock Bifurcation analysis and chaos control of the modified chua's circuit
  system.
\newblock \emph{Chaos Solit. Fractals}, 77, 332--339.

\end{thebibliography}
\end{document}